# A direct time-domain FEM modeling of broadband frequency-dependent absorption with the presence of matrix fractional power: Model I

W. Chen

Simula Research Laboratory, P. O. Box. 134, 1325 Lysaker, Norway

(21 April 2002)

## 1. Summary

The frequency-dependent attenuation of broadband acoustics is often confronted in many different areas. However, the related time domain simulation is rarely found in literature due to enormous technical difficulty. The currently popular relaxation models with the presence of convolution operation require some material parameters which are not readily available. In this study, three reports are contributed to address broadband ultrasound frequency-dependent absorptions using the readily available empirical parameters. This report is the first in series concerned with developing a direct time domain FEM formulation. The next two reports are about the "**frequency decomposition model**" and the "**fractional derivative model**".

In contrast to the traditional proportional damping model (e.g. Rayleigh damping), our FEM model is derived with a **modified mode superposition of broadband damping effect**, where the fractional power of a matrix is present to be fully consistent with empirical frequency-dependent attenuation law, in the cases of both single and broadband excitations. The constitutive analysis shows that the presented FEM modelling does not violate the principle of causality. The term "direct FEM" implies that the present FEM model can not be directly converted back to the corresponding time-domain PDE model since we applied a mode superposition assumption in the numerical modelling of damping effect.

The implementation is straightforward, but computing cost of matrix fractional power is a major concern. The related computational stability remains yet to discuss (probably just the same as for the Rayleigh model). Since the superposition principle does not hold in the nonlinear problems, the present methodologies can not be directly extended to the nonlinear dissipative ultrasound propagation. A combination of the present scheme and the linearization iteration [1] may be a solution. In addition, our model may be inductive to develop some **modal testing approach to determine the damping parameters**.

## 2. Frequency-dependent absorption

In numerous experiments of soft tissue ultrasound, the frequency-power attenuation due to the absorption and scattering is observed, i.e.

$$p(x + \Delta x) = p(x)e^{-\alpha(f)\Delta x}, \tag{1}$$

where $p$ represents the amplitude of acoustic pressure, and $\alpha(f)$ denotes the tissue-specific attenuation coefficient and is a power function of frequency within most useful frequency spectrum:

$$\alpha(f) = \alpha_0 f^y, \qquad y \in [0,2]. \tag{2}$$

Here $f$ is the frequency, and $\alpha_0$ the attenuation constant and $y$ the frequency-power exponent varying from 0 to 2. These tissue-dependent parameters in (2) are obtained by fitting experimental data with autocorrelation-based model approach.

### 3. Time-domain model for single frequency excitation

Assuming the viscous absorption depends solely on the velocity, the dissipative ultrasound wave model can be generally stated as

$$\frac{1}{c^2}\ddot{p} + z(\dot{p}) - \nabla^2 p = 0, \qquad (3)$$

where $z(\dot{p})$ is an implicit expression of viscous function, and upper dot denotes the temporal derivative. The FEM analogization of spatial Laplacian produces

$$\ddot{p} + c^2 z(\dot{p}) + c^2 Kp = g(t), \qquad (4)$$

where $K$ is the positive symmetric FEM interpolation matrix of self-adjoint Laplace operator, and vector $g(t)$ is the external excitation source due to the given boundary conditions. If the viscous term is further assumed to be linearly dependent on the velocity, (4) can be reduced to the standard damped wave equation

$$\ddot{p} + c^2 \gamma \dot{p} + c^2 Kp = g(t), \qquad (5)$$

where $\gamma$ is the viscous coefficient. In the case of a singular frequency excitation $g(t)$, ref. 1 derived

$$\ddot{p} + 2\alpha_0 c f^y \dot{p} + c^2 Kp = g(t), \qquad (6)$$

in corresponding to the empirical frequency-dependent attenuation (1). The model (6) is straightforward to embed the experimental parameter and very easy and efficient to implement. We can retrieve (6) back to a constitutive model of viscoelastic materials:

$$\begin{aligned}\sigma(t) &= E_0\varepsilon(t) + 2\alpha_0 c\rho f^{y}\dot{\varepsilon}(t) \\ &= E_0\varepsilon(t) + \beta f^{y}\dot{\varepsilon}(t)\end{aligned}, \quad (7)$$

where $\rho$ denotes material density; $E_0$ is the static elastic modulus; $\sigma$ and $\varepsilon$ are stress and strain, respectively. (7) underlies frequency dependence and requires three parameters. It is worth pointing out that (7) clearly does not violate the causality principle. Very recent theses 3 and 4 have a comprehensive survey on various existing damping models for elastoviscous materials. It is found that this model is distinct from all others in that it explicitly comprises of the empirical frequency-dependent attenuation formula (1).

Unfortunately, the explicit presence of frequency variable also causes the FEM time domain model (6) can not be simply extended to the case of a broadband excitation, which, however, is much more frequently confronted in ultrasound clinical practice. So, without further modifications, (6) has limited practical utility for medical ultrasound simulation. The superposition analysis and the frequency decomposition are two strategies to revise (6) to accommodate broadband frequency dependence. The next section focuses on the former approach, while we will address the latter strategy in the subsequent report II.

## 4. A modified mode superposition of broadband damping effect

The central goal of the time-domain mathematical modeling is to reflect any order (0-2) frequency-dependent absorption mechanism subject to a broadband excitation. This has widely been seen as a formidable task. With the help of a modified mode superposition assumption of broadband damping effect, this section develops a direct FEM formulation which accurately describes the energy attenuation behavior of dissipative broadband ultrasound propagation with the available empirical parameters. It is worth noting that the fractional power of a matrix appears in the resulting numerical model.

Without the explicit form $z(\dot{p})$ in the semi-discrete modelling (4), the total numerical damping effect matrix cannot be constructed from individual element damping matrix as did for the diffusion matrix $K$. Recalling our ultimate purpose is to analogize the overall energy dissipation during the propagation, it is reasonable to assume the orthogonal similarity of assembly FEM damping matrix to the diffusion matrix $K$ and the linear dependence of damping on velocity. Thus, the natural frequency-dependent damping is decomposed as

$$\phi_i^T z(\dot{p})\phi_j = s(\omega_i)\delta_{ij}\dot{q}, \tag{8}$$

where $q_i=\phi_i^T p$ is modal coordinates, $\delta_{ij}$ is the Kronecker delta, and $\omega_i^2$ and $\phi_i$ are respectively real eigenvalues and orthogonal eigenvectors of the i-th mode,

$$\sum_i \omega_i^2 = \Phi K \Phi^T. \tag{9}$$

(8) means the assumption that the total damping in the ultrasound propagation is the sum of individual damping in each eigenmode. $\omega_i^2$ account for the natural circular frequency components here of i-th mode. It is stressed that (8) was assumed with implicit equation (4) rather than with the standard explicit damped equation (5). In terms of (8), a mode matrix transform reduces (4) to a set of uncoupled ordinary differential equations of the form

$$\ddot{q}_i + c^2 s(\omega_i)\dot{q}_i + c^2 \omega_i^2 q_i = \hat{g}_i(t), \tag{10}$$

where $\hat{g}_i(t) = \phi_i^T g_i(t)$ is the associated modal force. (10) is the equilibrium equation governing motion of each single degree of freedom system. It is crucially important that

(8) does not necessarily represent the proportional damping of the standard superposition analysis [5]:

$$s(\omega_i) = 2\xi_i \omega_i \delta_{ij}, \qquad (11)$$

where $\xi_i$ are modal damping ratio parameters. In terms of (11), the known Rayleigh proportional damping model is given by

$$\ddot{p} + c^2(\alpha I + \beta K)\dot{p} + c^2 K p = g(t), \qquad (12)$$

where $I$ is the unit matrix; constants $\alpha$ and $\beta$ are calculated by two given damping ratios that correspond to two unequal frequencies. Consequently, the Rayleigh damping characterizes only zero ($\beta=0$) or square ($\alpha=0$) frequency dependence. By using a trial-error approach adapting the parameter $\alpha$ and $\beta$ of Rayleigh damping model, Wojcik et al. [1] tried to reflect arbitrary frequency dependence. Their strategy is very tricky and tedious.

In terms of (10), the damping in one eigenmode could be observed by imposing initial or boundary conditions corresponding to that mode only and measuring the amplitude decay [5]. It is exactly in this way that we get the empirical frequency-dependent absorption (1), which is found widely effective with biomaterials. According to the frequency-dependent viscous coefficient formula under a single frequency excitation [2] (also see the previous (6)), we have

$$s(\omega_i) = \frac{2\alpha_0 \omega_i^y}{c}. \qquad (13)$$

With the damping mode superposition (13), the model (10) is restated as

$$\ddot{q}_i + 2\alpha_0 c \omega_i^y \dot{q}_i + c^2 \omega_i^2 q_i = \hat{g}_i(t). \qquad (14)$$

The above (14) is further converted back to a matrix form via the inverse orthogonal mode matrix transform, i.e.

$$\ddot{p} + 2\alpha_0 c K^{y/2} \dot{p} + c^2 K p = g(t). \tag{15}$$

It is obvious that (15) readily takes into account of frequency-dependent viscous effects of a multitude of frequency components with realistic empirical coefficients $\alpha_0$ and $y$ provided that the mode superposition assumption (13) is satisfied.

The empirical formula (2) of frequency dependent attenuation is also often expressed as

$$\alpha(f) = \alpha_1 + \alpha_0 f^y, \quad y \in [0,2]. \tag{16}$$

In terms of the present modified mode superposition model of damping effect, we have the corresponding time domain model

$$\ddot{p} + 2c(\alpha_1 I + \alpha_0 K^{y/2})\dot{p} + c^2 K p = g(t), \tag{17}$$

We need to stress that the frequency $f$ of excitation source $g(t)$ in (1), (2), (6) and (7) are different from the system natural circle frequency $\omega$ in (13) and (14). Since the natural frequency spectrum $\omega$ of large FEM discretization matrix associated with our ultrasound simulation has to be broad enough for multiscale broadband excitation, the $\omega$ spectrum covers the external broadband spectrum $f$. Otherwise, we could not get the accurate FEM solution. This also implies that the FEM simulation of a broadband excitation is a multiscale problem and could be much more expensive than that of a single frequency excitation.

It is observed that both the present (15) and (17) and the Rayleigh model (12) can not be traced back completely to the original PDE model (3) since the model superposition assumption (8) was made with the implicit damping term $z(\dot{p})$ in the discretized FEM formulation (4) rather than with the explicit $c^2\gamma\dot{p}$ in (5). Therefore, (15) and (17) are the mixed models, in the sense that the viscous effect is modeled via a combination of numerical superposition assumption (13) and a linear dependence on velocity, while the diffusion and inertia terms are fully described by differential operators. This is the essence of the superposition analysis of the damping, where we do not need to calculate and assembly the damping matrix, but only the diffusion (stiffness and/or mass) matrix. It must be stressed that (15) and (17) are only a computational way of numerically representing broadband damping and does not imply all complicated physical or chemical mechanisms for damping.

### 5. Related equations and dispersion analysis

In short, (13), (15) and (17) are essential assumption and result of the present direct FEM model, respectively, which represents a dependence of the viscous effect on a broad frequency spectrum. The Rayleigh damping model is a special case of our model when $y=0$ or 2. The propagation of sound through a viscous fluid is also governed by the augmented wave equation

$$\nabla^2 p = \frac{1}{c^2}\frac{\partial^2 p}{\partial t^2} - \frac{4\gamma}{3\rho c^2}\frac{\partial}{\partial t}\nabla^2 p, \qquad (18)$$

where $\gamma$ is still a positive viscous coefficient and $\rho$ denotes the ambient density. The equation describes both dispersion (waveform alternation) and attenuation behavior. The FEM discretization of (18) yields

$$\ddot{p} + \frac{4\gamma}{3\rho}K\dot{p} + c^2 Kp = g(t). \qquad (19)$$

Compared (15) and (19), it is observed that when $y=2$, our model (15) brings out the square frequency dependence as does the augmented wave model (19), but two have different forms of damping coefficient constant.

With the Duhamel integral and when $\alpha_0\omega_i^y<1$, we can have the solution of each single modal equation (14)

$$q_i(t) = \frac{1}{\hat{c}_i}\int_0^t \hat{g}(\tau)e^{-\alpha_0 c\omega_i^y(t-\tau)}\sin\hat{c}_i(t-\tau)d\tau + e^{-\alpha_0 c\omega_i^y t}\{\alpha_i\sin\hat{c}_i t + \beta_i\cos\hat{c}_i t\} \qquad (20)$$

where $\alpha_i$ and $\beta_i$ are calculated using the initial conditions, and the distorted phase velocity $\hat{c}_i$ dependent on the attenuation is calculated by

$$\hat{c}_i = \omega_i c\sqrt{1-\alpha_0^2\omega_i^{2y-2}}. \qquad (21)$$

(21) shows that the damped phase speed $\hat{c}_i$ increases with the frequency due to the absorption and is lower than the lossless speed, and the viscous effect causes the dispersion. (20) illustrates how the group waveform is distorted on each frequency component. The second term of (20) is the transient response which decays with time, while the first term is the steady response. In addition to the above oscillatory case, the solutions are divided into overdamped ($\alpha_0\omega_i^y<1$)

$$q_i(t) = \int_0^t \hat{g}(\tau)h(t-\tau)d\tau + \alpha_i e^{-\alpha_0 c\omega_i^y t - c\omega_i t\sqrt{\alpha_0^2\omega_i^{2y-2}-1}} + \beta_i e^{-\alpha_0 c\omega_i^y t + c\omega_i t\sqrt{\alpha_0^2\omega_i^{2y-2}-1}} \qquad (22)$$

and critically damped ($\alpha_0\omega_i^y=1$)

$$q_i(t) = \int_0^t \hat{g}(\tau)r(t-\tau)d\tau + e^{-\alpha_0 c\omega_i^y t}(\alpha_i + \beta_i t), \qquad (23)$$

where $h$ and $r$ are the impulse response function of systems.

The resulting solution can be simply obtained by a modal matrix transform, i.e., $p=\Phi q$. Compared with the other models [6], our model is mathematically much easier to understand and implement.

## 6. Computing aspects

With the present model (15), the calculation of the fractional power of a matrix needs to be address. The orthodox analytical approach is costly singular value decomposition, i.e.

$$K^{y/2} = \Phi^T \left(\sum \omega_i^2\right)^{y/2} \Phi. \tag{24}$$

On the other hand, the numerical computation requires $O(n^3)$ operations [7]. The popular methods are the Schur decomposition, Pade approximation, and iterative method. By far, we have no idea which of these algorithms is the most parallel friendly. The handling of the matrix power may be the major weakness of the present model. The frequency decomposition model in the subsequent report will circumvent this costly task with a reasonable compromise in the accuracy.